\documentclass[american,aps,pra,reprint,longbibliography,superscriptaddress]{revtex4-2}
\usepackage{lmodern}
\usepackage{lmodern}
\usepackage[T1]{fontenc}
\usepackage[utf8]{inputenc}
\usepackage{color}
\usepackage{babel}
\usepackage{units}
\usepackage{amsmath}
\usepackage{amssymb}
\usepackage{graphicx}
\usepackage[normalem]{ulem}
\usepackage[pdfusetitle,
 bookmarks=true,bookmarksnumbered=false,bookmarksopen=false,
 breaklinks=false,pdfborder={0 0 1},backref=false,colorlinks=true]
 {hyperref}
\hypersetup{
 pdfborderstyle=,citecolor=blue,linkcolor=teal,urlcolor=blue}

\makeatletter

\usepackage{xcolor}
\usepackage{babel}
\usepackage{units}

\makeatother

\begin{document}
\global\long\def\i{\mathrm{i}}%
\global\long\def\d{\mathrm{d}}%
\global\long\def\e{\mathrm{e}}%
\global\long\def\Tr{\mathrm{Tr}}%
\global\long\def\bra#1{\langle#1|}%
\global\long\def\ket#1{|#1\rangle}%
\global\long\def\braket#1#2{\left\langle #1|#2\right\rangle }%
\global\long\def\ketbra#1#2{\left|#1\right\rangle \!\left\langle #2\right|}%
\global\long\def\dbra#1{\langle\!\langle#1|}%
\global\long\def\dket#1{|#1\rangle\!\rangle}%
\global\long\def\dbraket#1#2{\left\langle \!\left\langle #1|#2\right\rangle \!\right\rangle }%
\global\long\def\dketbra#1#2{\left.\!\left|#1\right\rangle \!\right\rangle \!\left\langle \!\left\langle #2\right|\!\right.}%

\title{Optical creation of dark--bright soliton lattices in one-dimensional multicomponent Bose--Einstein condensates}
\author{Y.~Braver}
\affiliation{Institute of Theoretical Physics and Astronomy, Faculty of Physics,
Vilnius University, Saulėtekio 3, LT-10257 Vilnius, Lithuania}

\author{D.~Burba}
\affiliation{Institute of Theoretical Physics and Astronomy, Faculty of Physics,
Vilnius University, Saulėtekio 3, LT-10257 Vilnius, Lithuania}

\author{Th.~Busch}
\affiliation{Quantum Systems Unit, Okinawa Institute of Science and Technology
Graduate University, Okinawa 904-0495, Japan}

\author{G.~Juzeli{\=u}nas}
\affiliation{Institute of Theoretical Physics and Astronomy, Faculty of Physics,
Vilnius University, Saulėtekio 3, LT-10257 Vilnius, Lithuania}

\author{P.~G.~Kevrekidis}
\affiliation{Department of Mathematics and Statistics, University of Massachusetts Amherst, Amherst, MA 01003, USA}

\affiliation{Department of Physics, University of Massachusetts Amherst, Amherst, MA 01003, USA}

\affiliation{Department of Mechanical Engineering, Seoul National University, 1 Gwanak-ro, Gwanak-gu, Seoul 08826, South Korea}

\begin{abstract}
We present a widely accessible and experimentally realizable technique for the controlled creation of dark--bright solitons and soliton lattices in one-dimensional atomic Bose--Einstein condensates. The method is based on preparing the condensate in a dark state of a $\Lambda$-coupled three-level system. Numerical simulations of the corresponding two-component system reveal that individual dark--bright solitons created through this scheme can survive over experimentally accessible timescales, even when the coupling laser fields are switched off. Meanwhile, the fate of soliton lattices upon the quench of the fields depends on the scattering lengths. When they are all equal, the lattice is found to persist on timescales comparable to the condensate lifetime, even though the analysis of dynamical stability reveals that they possess unstable modes. In this case the resulting destabilization is not found to be detrimental, as it leads to recurrent dynamics. However, for unequal scattering lengths the lattice structure gets destroyed once the instability sets in, which happens after a few tens of milliseconds after the quench of the optical fields.

\end{abstract}
\maketitle

\section{Introduction}

Atomic Bose–Einstein condensates (BECs) constitute a canonical platform for the study of nonlinear coherent structures due to the BECs' high degree of controllability, long coherence times, and potential multicomponent and multidimensional nature \cite{pitaevskiiBoseEinsteinCondensationSuperfluidity2016}. In single-component settings, they support dark~\cite{frantzeskakisDarkSolitonsAtomic2010} and bright~\cite{abdullaevDYNAMICSBRIGHTMATTER2012} matter-wave solitons in the presence of repulsive and attractive interactions, respectively. These excitations have been experimentally realized in elongated condensates~\cite{kevrekidisEmergentNonlinearPhenomena2008}, starting with the early observations of dark solitons \cite{burger1999dark,denschlag2000generating} and bright solitary waves \cite{khaykovich2002formation,strecker2002formation}.

Multicomponent condensates enable a richer class of nonlinear states. In particular, two-component BECs support dark--bright (DB) solitons~\cite{Busch2001}, where a bright component is localized within the density dip of a dark soliton \cite{beckerOscillationsInteractionsDark2008,middelkampDynamicsDarkBright2011,hamnerGenerationDarkBrightSoliton2011,Rabec2025}. Despite their featuring bright solitonic structures, these can arise even in repulsive systems through an effective potential induced by the dark soliton, a mechanism with direct analogies in nonlinear optics \cite{trilloOpticalSolitaryWaves1988,christodoulidesBlackWhiteVector1988,ostrovskayaNonlinearTheorySolitoninduced1998} and connections to integrable vector nonlinear Schr{\"o}dinger models \cite{Sheppard1997}. Extensions to spinor condensates further broaden the spectrum of possible soliton complexes \cite{nistazakisBrightdarkSolitonComplexes2008,xiongDynamicalCreationComplex2010,bersanoThreeComponentSolitonStates2018,changObservationSpinorDynamics2004,changCoherentSpinorDynamics2005,Kawaguchi2012,stamper-kurnSpinorBoseGases2013,iedaMatterWaveSolitonsSpinor2004,romero-rosControlledGenerationDarkbright2019} and enable fundamental insights
into their interactions~\cite{lannigCollisionsThreeComponentVector2020}.

A key property of solitons is their particle-like character, enabling effective descriptions of their motion in external potentials and of their interactions \cite{frantzeskakisDarkSolitonsAtomic2010,kevrekidisSolitonsCoupledNonlinear2016,katsimigaDarkbrightSolitonInteractions2017,yanMultipleDarkbrightSolitons2011,Meng_24}. These features have been extensively explored experimentally, including collisions of dark and bright solitons \cite{wellerExperimentalObservationOscillating2008,stellmerCollisionsDarkSolitons2008,nguyenCollisionsMatterwaveSolitons2014} and the controlled realization of different 
multicomponent (including three-component ones) and magnetic solitonic excitations \cite{beckerOscillationsInteractionsDark2008,farolfiObservationMagneticSolitons2020,bersanoThreeComponentSolitonStates2018}. Recent experiments have further advanced control over such nonlinear structures, including the interferometric manipulation of bright solitons \cite{wales2020splitting}.
Current efforts increasingly focus on multisoliton configurations, where collective effects emerge. In this direction, dense ensembles of DB solitons, which are sometimes described as soliton gases, have  been recently observed \cite{mossman2024observation}, opening new avenues for studying nonequilibrium dynamics and emergent behavior in multicomponent nonlinear quantum systems.

The aim of the present work is to describe and characterize a method to generate multicomponent solitary waves, notably
dark--bright solitons and lattices thereof, via the preparation of a BEC
in a dark state of a $\Lambda$-coupled three-level system \citep{Juzeliunas2005,Braver2025,Hamedi2022}. We note that such
a technique is in line with the significant current interest in the 
experimental production of such configurations. Among recent examples,
we note that our technique is more controllable than the one experimentally demonstrated based on dynamical instabilities in counterflow dynamics, which is less predictable in its coherent
structure generation~\cite{hamnerGenerationDarkBrightSoliton2011}. It is arguably more comparable to and similarly versatile
as the winding technique of~\cite{mossman2024observation}. Another related proposal of the creation of dark--bright solitons was considered in Ref.~\cite{Juzeliunas2007} using the tripod scheme without, however, direct extensions to soliton lattices. This motivates
our exploration of creating DB solitons and their lattices both within
the integrable Manakov limit~\cite{manakov1974}, where such multisoliton solutions are known to exist~\cite{Sheppard1997,Yan2015}, as well as for the situation where scattering lengths are non-equal, as is the case for $^{87}$Rb. In what follows, these two cases are investigated  in the
absence and in the presence of an optical-box confinement.

\section{Model\label{sec:The-model}}

\begin{figure}
\begin{centering}
\includegraphics{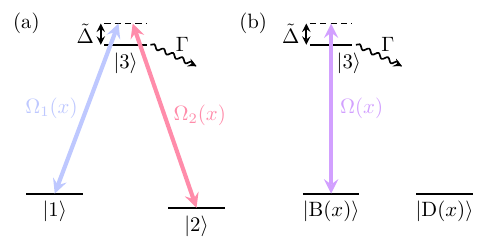}
\par\end{centering}
\caption{The $\Lambda$ atom--light coupling scheme. (a) Bare-state basis $\{|1\rangle, |2\rangle, |3\rangle\}$: states $|1\rangle$ and $|2\rangle$ of the ground state multiplet are coupled to an excited state $|3\rangle$ via position-dependent Rabi frequencies $\Omega_1(x)$ and $\Omega_2(x)$, with single-photon detuning $\tilde{\Delta}$ and spontaneous emission rate $\Gamma$. (b) Dark--bright basis $\{|\mathrm{B}(x)\rangle, |\mathrm{D}(x)\rangle, |3\rangle\}$: the bright state $|\mathrm{B}(x)\rangle$ couples to $|3\rangle$ with total Rabi frequency $\Omega(x) = \sqrt{|\Omega_1(x)|^2 + |\Omega_2(x)|^2}$, while the dark state $|\mathrm{D}(x)\rangle$ is decoupled from the light fields.
\label{fig:lambda-scheme}
}
\end{figure}

We consider a three-level atomic system coupled in a $\Lambda$ configuration, as shown in Fig.~\ref{fig:lambda-scheme}(a). The single-particle Hamiltonian is given by
\begin{equation}
\hat{H}(x)=-\partial^2_{x}+\hat{H}_{0}(x).
\end{equation}
The second term describes the internal degrees of freedom and can be written in the rotating frame as
\begin{equation}
\hat{H}_{0}(x)=-\i\tfrac{\Gamma}{2}\ketbra 33+[\Omega_{1}(x)\ketbra 31+\Omega_{2}(x)\ketbra 32+{\rm H.c.}].\label{eq:H0}
\end{equation}
For simplicity, we consider the case where the single- and two-photon detunings are zero. However, the dark-state analysis remains valid over some finite interval of detunings \citep{Braver2025}. Furthermore, we have introduced the spontaneous decay rate
$\Gamma$ for the excited state $\ket 3$ represented through the
imaginary part of the excited-state energy. Following Refs.~\citep{Lacki2016,Jendrzejewski2016,WangY2018}, to create the subwavelength barriers we choose the Rabi frequencies of the coupling beams to be
\begin{equation}
\begin{split}\Omega_{1}(x) & =\Omega_{10}\sin x,\\
\Omega_{2}(x) & =\Omega_{20}
\end{split}
\end{equation}
with $\Omega_{10} \gg \Omega_{20}$, where the amplitudes $\Omega_{10}$ and $\Omega_{20}$ are taken to be  real and positive. Throughout this work, we use recoil units, where the length is measured in units of $1/k=\lambda_{\rm L}/2\pi$, with $\lambda_{\rm L}$ being the wavelength of the laser field characterized by $\Omega_{1}$. The energy is measured in recoil energies $E_{{\rm R}}=\hbar^{2}k^{2}/2m$, where $m$ is the mass of the atom.

The Hamiltonian in Eq.~\eqref{eq:H0} possesses a dark eigenstate
given by 
\begin{equation}
\ket{{\rm D}(x)}=\frac{1}{\sqrt{1+|\zeta(x)|^{2}}}[\ket 1-\zeta(x)\ket 2],
\end{equation}
where 
\begin{equation}
\zeta(x)=\frac{\Omega_{1}(x)}{\Omega_{2}(x)}.\label{eq:zeta}
\end{equation}
The dark state has no contribution from the lossy excited state $\ket 3$,
ensuring an infinite lifetime.
The remaining two eigenstates of this Hamiltonian are given (for the simpler case of $\Gamma=0$) by
\begin{equation}
\ket{\pm}=\frac{1}{\sqrt{2}}\left(\ket{{\rm B}}\pm\ket 3\right),\quad\mathrm{where}\quad\ket{{\rm B}}=\frac{\zeta^{*}\ket 1+\ket 2}{\sqrt{1+|\zeta|^{2}}},\label{eq:pm-states}
\end{equation}
with 
\begin{equation}
\hat{H}_{0}\ket{\pm}=\pm\Omega\ket{\pm}\quad\mathrm{and}\quad\Omega=\sqrt{\left|\Omega_{1}\right|^{2}+\left|\Omega_{2}\right|^{2}}.\label{eq:pm-eigenvalue_eq}
\end{equation}
In Eq.~(\ref{eq:pm-states}), $\ket{{\rm B}}$ is the so-called bright
state, representing a superposition of atomic ground states orthogonal
to the dark state $\ket{{\rm D}}$. Later we will consider the adiabatic
motion of atoms in the dark-state manifold, which is justified when the
total Rabi frequency $\Omega$ is much larger than the characteristic
kinetic energy of the atomic center of mass motion.

Now let us consider the equations for an atomic BEC interacting with
the laser fields. We start with the Schrödinger equation for the full
state vector $\ket{\Phi(t,x)}$ of a single atom 
\begin{equation}
\i\partial_t\ket{\Phi(t,x)}=\hat{H}(x)\ket{\Phi(t,x)}\label{Schroedinger-eq}
\end{equation}
and use the expansion 
\begin{equation}
\ket{\Phi(t,x)}=\sum_{i=1}^{3}\Phi_{i}(t,x)\ket i.\label{eq:Phi-state-vector}
\end{equation}
Here $\ket i$ are the bare atomic states 
of $\hat{H}_{0}$ appearing in Eq.~(\ref{eq:H0}). To account for the interaction
between the atoms, we use the Gross--Pitaevskii approach for the
multicomponent BECs \citep{pitaevskiiBoseEinsteinCondensationSuperfluidity2016}. In practice, this
amounts to supplementing the Schrödinger equations for $\Phi_{i}(x)$
with the nonlinear terms, thereby promoting the single-particle 
wave functions to the wave functions (order parameters) of the components
of the condensate \citep{Juzeliunas2005}. This way we arrive at

\begin{align}
\i\partial_t\Phi_{1} & =\left(-\partial^2_{x}+g_{11}|\Phi_{1}|^{2}+g_{12}|\Phi_{2}|^{2}\right)\Phi_{1}+\Omega_{1}^{*}\Phi_{3},\nonumber \\
\i\partial_t\Phi_{2} & =\left(-\partial^2_{x}+g_{12}|\Phi_{1}|^{2}+g_{22}|\Phi_{2}|^{2}\right)\Phi_{2}+\Omega_{2}^{*}\Phi_{3},\nonumber \\
\i\partial_t\Phi_{3} & =\left(-\partial^2_{x}-\i\frac{\Gamma}{2}\right)\Phi_{3}+\Omega_{1}\Phi_{1}+\Omega_{2}\Phi_{2}.\label{eq:full-gpe}
\end{align}
For atoms adiabatically following the dark state, the population of
the excited state $\ket 3$ described by the wavefunction $\Phi_{3}$
is small at all times, allowing us to neglect collisions 
in the equation for $\Phi_{3}$. The coefficients $g_{ij}$ describing
interaction between the atoms in the corresponding internal states
are related to the $s$-wave scattering lengths $a_{ij}$ as $g_{ij}=8a_{ij}N/kR^{2}$
where $R$ is the radius of the cross-section of the cloud \cite{Juzeliunas2007}. We therefore absorb
the total number of particles $N$ in the coupling strengths and
normalize the wave function to unity: $\sum_{i=1}^{3}\int\d x\,|\Phi_{i}|^{2}=1$.
Under the constraint of a fixed total number of particles, Eqs.~\eqref{eq:full-gpe}
then possess a stationary state of the form $\Phi_{i}(t,x)=\varphi_{i}(x)\e^{-\i\mu t}$, where $\mu$ is the chemical potential \cite{Leggett2001}.

To obtain an equation for the evolution of atoms adiabatically following
the dark state, we perform the following change of variables:
\begin{equation}
\begin{split}\Phi_{{\rm D}} & =\frac{1}{\sqrt{1+|\zeta|^{2}}}(\Phi_{1}-\zeta^{*}\Phi_{2}),\\
\Phi_{{\rm B}} & =\frac{1}{\sqrt{1+|\zeta|^{2}}}(\zeta\Phi_{1}+\Phi_{2}).
\end{split}
\label{eq:PhiD-PhiB}
\end{equation}
Inserting these definitions into Eqs.~(\ref{eq:full-gpe}), we obtain
a set of equations for $\Phi_{{\rm D}}$, $\Phi_{{\rm B}}$ and $\Phi_{3}$,
in which the wave functions $\Phi_{{\rm D}}$
and $\Phi_{{\rm B}}$ are coupled via non-adiabatic terms. Under the
adiabatic assumption of slow center-of-mass motion for atoms in the
dark internal states, justified for the systems to be considered,
and the assumption of equal nonlinear couplings ($g_{11}=g_{12}=g_{22}=g$),
we reach the dark-state Gross--Pitaevskii equation \citep{Juzeliunas2005}
\begin{equation}
\i\partial_t\Phi_{{\rm D}}=-\partial^2_{x}\Phi_{{\rm D}}+U\Phi_{{\rm D}}+g|\Phi_{{\rm D}}|^{2}\Phi_{{\rm D}}.\label{eq:ds-gpe}
\end{equation}
This equation is fully decoupled from those for $\Phi_{{\rm B}}$ and $\Phi_{3}$, mirroring (in the mean-field context) the decoupling of single-particle dark state $\ket{{\rm D}}$ from states $\ket{{\rm B}}$ and $\ket{3}$ [cf. 
Fig.~\ref{fig:lambda-scheme}(b)].
In contrast, the equation for the “bright-state BEC” $\Phi_{{\rm B}}$
remains coupled to $\Phi_{3}$, allowing atoms in $\Phi_{{\rm B}}$ to
transition into $\ket 3$ and decay, which is also displayed in Fig.~\ref{fig:lambda-scheme}(b).

It is apparent from Eq.~(\ref{eq:ds-gpe}) that reduction to the
dark-state manifold results in the appearance of a geometric scalar potential
$U(x)$, which emerges from the position-dependence of the atom--light
coupling \citep{Dalibard2011,Goldman2014} and for the present scheme is given by  \citep{Juzeliunas2005EIT,Juzeliunas2005,Lacki2016}
\begin{equation}
U=\frac{(\partial_x\zeta^{*})(\partial_x\zeta)}{(1+|\zeta|^{2})^{2}}=\frac{\epsilon^{2}\cos^{2}x}{(\epsilon^{2}+\sin^{2}x)^{2}},\label{eq:U}
\end{equation}
where $\epsilon = \Omega_{20}/\Omega_{10}$. The subwavelength barriers are created for $\epsilon\ll 1$ \citep{Lacki2016,Jendrzejewski2016,WangY2018}.

Once the solution to Eq.~(\ref{eq:ds-gpe}) is found, one can return
to the wave functions $\Phi_{1}$ and $\Phi_{2}$ of the internal-state
basis according to

\begin{equation}
\Phi_{1}=\frac{1}{\sqrt{1+|\zeta|^{2}}}\Phi_{{\rm D}},\qquad\Phi_{2}=-\frac{\zeta}{\sqrt{1+|\zeta|^{2}}}\Phi_{{\rm D}}.
\label{eq:ds-comps}
\end{equation}
This follows from Eqs.~(\ref{eq:PhiD-PhiB}) with $\Phi_{{\rm B}}=0$
since for adiabatic atomic motion in the dark state, described by Eq.~(\ref{eq:ds-gpe})
for the wavefunction $\Phi_{{\rm D}}$, the system is almost completely
in the dark state with a negligible population of the bright and excited
states.

In our simulations, the states $\ket 1$ and $\ket 2$ are taken to
be two hyperfine levels of $^{87}{\rm Rb}$: $\ket 1=\ket{F=1,m_{F}=-1}$
and $\ket 2=\ket{F=2,m_{F}=1}$, while state $\ket 3$ is the $5\,{}^{2}{\rm P}_{\frac{3}{2}}$
state, whose decay rate is $\Gamma_*=\unit[38.1]{MHz}$ \citep{Gutterres2013}. Due to the decay, the energies and chemical potentials of the stationary states acquire an imaginary part.
For the dark stationary states considered in this work, these imaginary parts turn out to be
of order $10^{-5}$, leading to the loss of 62\% of atoms in 1 second. However, this loss can only take place as long as the optical fields are on. There is no loss during the dynamics ensuing after the quench of the fields, which is the scenario that we will focus on. Notably, we found that none of the reported results depend on the values of the decay rate (we also checked larger values, up to $100\Gamma_*$). Therefore, we present calculations obtained for $\Gamma=0$, allowing us to consider real stationary states.

The scattering lengths of the considered states are \citep{Egorov2013}: $a_{11}=100.4a_{0}$,
$a_{12}=98.006a_{0}$, $a_{22}=95.44a_{0}$, where $a_{0}$ is the Bohr
radius. In dimensionless units, we set $g_{11}=100.4$ (with $g_{12}$ and $g_{22}$ following accordingly) corresponding to a radius of cross-section
of the order of $R\sim\unit[2]{\mu m}$ and the number of atoms being
of the order of $N\sim10^{4}$. We assume a laser wavelength of
$\lambda_{\rm L}=\unit[780]{nm}$, the amplitude of the Rabi frequency $\Omega_{2}$ is chosen to be $\Omega_{20}=2000$, and we set $\epsilon=0.1$.
Unless stated otherwise, we use these values of parameters for all our calculations. When solving Eq.~\eqref{eq:ds-gpe}, we use $g=g_{11}$.

\section{Single dark--bright soliton}

\begin{figure}
\begin{centering}
\includegraphics{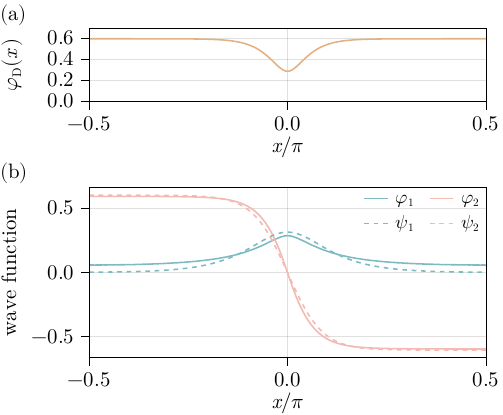}
\par\end{centering}
\caption{Stationary state of the system interacting with the light fields for $x\in[-\pi/2, \pi/2]$.
Nonlinear coupling ratios corresponding to \textsuperscript{87}Rb were used (see Section \ref{sec:The-model}).
(a) Lowest-energy dark-state wave function $\varphi_{{\rm D}}$ as the stationary solution of Eq.~\eqref{eq:ds-gpe} with $g=g_{11}$. (b) Solid lines: stationary solution of Eqs.~\eqref{eq:full-gpe} representing the lowest-energy state of the dark-state manifold. The wave function $\varphi_3$ is not shown since the population of this component is less than $10^{-5}$.
Dashed lines: corresponding stationary solution of Eqs.~\eqref{eq:free}.}\label{fig:phi_vs_psi_0_5pi}
\end{figure}

To get a basic understanding of the ground state of the system coupled
to the light fields, we start by considering the ground state of Eq.~\eqref{eq:ds-gpe}
for $x\in[-\pi/2,\pi/2]$, thereby taking into account only a single
peak of the scalar potential. The spatial part of the ground state
$\Phi_{{\rm D}}(t,x)=\varphi_{{\rm D}}(x)\e^{-\i\mu t}$, calculated
under the Neumann boundary conditions and setting $g=g_{11}$,
is depicted in Fig.~\ref{fig:phi_vs_psi_0_5pi}(a). The wavefunction
has a dip of the density at the position of the peak of $U$, but, being the
ground state, features no nodes. In terms of components, it follows
from Eq.~\eqref{eq:ds-comps} and $\zeta=\epsilon^{-1}\sin x$ that
$\varphi_{2}$ mostly inherits the shape of $\varphi_{{\rm D}}$,
but gets zeroed-out at the position of the potential peak, and changes
sign there. As a result, $\varphi_{2}$ acquires the shape of a dark
soliton. Meanwhile, $\varphi_{1}$ has a maximum at the position of
the peak of $U$ and decays away from it.
This results in $\varphi_{1}$ having the shape of a bright soliton,
although its tails do not decay exponentially.
These findings are shown in Fig.~\ref{fig:phi_vs_psi_0_5pi}(b),
with the curves $\varphi_{1}$ and $\varphi_{2}$ obtained by numerically
calculating the stationary state of Eqs.~\eqref{eq:full-gpe}. The
calculation is performed respecting the fact that $g_{ij}$ are
different, although in that case the resulting wave functions differ marginally
from the case of equal $g_{ij}$.

We are especially interested in the stability of the free solitons
when there are no light fields. To this end we consider quenching
the fields off (immediately once the state has been prepared, so that no significant loss of atoms can take place) and tracking the subsequent evolution governed by the {\it free}
equations
\begin{equation}
\begin{split}\i\partial_t\Psi_{1} & =\left(-\partial^2_x+g_{11}|\Psi_{1}|^{2}+g_{12}|\Psi_{2}|^{2}\right)\Psi_{1},\\
\i\partial_t\Psi_{2} & =\left(-\partial^2_x+g_{12}|\Psi_{1}|^{2}+g_{22}|\Psi_{2}|^{2}\right)\Psi_{2}.
\end{split}
\label{eq:free}
\end{equation}
It is well known that in the case of equal nonlinear couplings such
a system of equations possesses stationary solutions $\Psi_{i}(t,x)=\psi_{i}(x)\e^{-\i\mu_{i}t}$
of the form of dark--bright solitons \citep{Sheppard1997,Busch2001}.
Such solutions can be found numerically even away from the case of
equal $g_{ij}$. It is instructive to compare the stationary
states $\varphi$ of the system with the light fields on with the
stationary states $\psi$ with the fields off. These are obtained using the Newton--Raphson iteration with the states $\varphi_{1}$ and $\varphi_{2}$ as the
initial guess and under the constraint that the number of particles
in each component be preserved:
\begin{equation}
\int\d x\,|\varphi_{i}|^{2}=\int\d x\,|\psi_{i}|^{2}.\label{eq:constraint}
\end{equation}
The chemical potentials of the two components are therefore not fixed
but rather obtained during the iteration procedure. The comparison
is shown in Fig.~\ref{fig:phi_vs_psi_0_5pi}(b). It is apparent that
the shape of the dark soliton $\varphi_{2}$ closely matches that
of the stationary one ($\psi_{2}$). However, the match
between $\varphi_{1}$ and $\psi_{1}$ is less accurate due to the trigonometric-function-dictated decay of $\varphi_{1}$, reflected in the factor $1/\sqrt{1+|\zeta|^{2}}$.
To quantify the match of the wave functions, we calculate the overlap,
obtaining the value $\braket{\varphi}{\psi}=\sum_{i=1}^{2}\int\d x\,\varphi^{*}_{i}\psi_{i}=0.9964$.
Such a high value of the overlap is guaranteed by the fact that most
of the atoms (95\%) occupy state $\varphi_{2}$, which indeed matches
$\psi_{2}$ with high accuracy.

\begin{figure}
\begin{centering}
\includegraphics{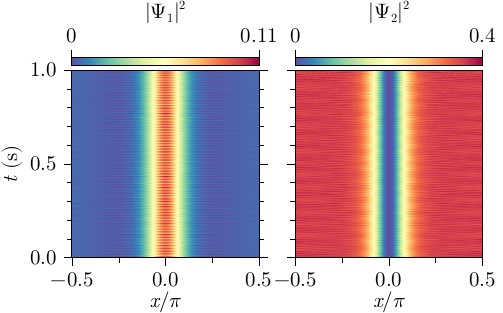}
\par\end{centering}
\caption{Evolution of
the densities $|\Psi_{1}|^2$ and $|\Psi_{2}|^2$
[governed by Eqs.~\eqref{eq:free}] starting from $\varphi_{1}$ and $\varphi_{2}$ --- the stationary state of the system interacting with the light fields, shown in Fig.~\ref{fig:phi_vs_psi_0_5pi}(b).
Nonlinear coupling ratios corresponding to \textsuperscript{87}Rb were used.}\label{fig:dynamics_0_5pi}
\end{figure}

In the case of equal nonlinear couplings, the dark--bright soliton
is dynamically stable~\cite{yang2010,KeverkidisBook}. To investigate the stability of the stationary state
$\psi$ obtained for unequal $g_{ij}$, we perform the Bogoliubov--de~Gennes (BdG) stability analysis~\cite{kevrekidisSolitonsCoupledNonlinear2016} by considering the evolution of the state
\begin{equation}
\Psi_{i}(t,x)=\psi_{i}(x)\e^{-\i\mu_{i}t}+[a_{i}(x)\e^{\lambda t}+b_{i}^{*}(x)\e^{\lambda^{*}t}]\e^{-\i\mu_{i}t},\label{eq:perturb}
\end{equation}
where $a_{i}$ and $b_{i}$ represent the perturbation. Inserting
this expression into the system \eqref{eq:free}, keeping only the linear
terms, and requiring the existence of a nontrivial solution leads
to the eigenvalue equation
\begin{equation}
\hat{G}(x)B(x)=\i\lambda B(x).\label{eq:BdG}
\end{equation}
Here $B(x)=\begin{pmatrix}a_{1}(x) & b_{1}(x) & a_{2}(x) & b_{2}(x)\end{pmatrix}^{\mathsf{T}}$
while $\hat{G}(x)$ is a 4-by-4 operator matrix containing the Laplace term and depending
on $\psi_{i}(x)$. It follows from Eq.~\eqref{eq:perturb} that eigenvalues
$\lambda$ having positive real part indicate an exponential growth
of the corresponding perturbation.

For the state $\psi$ shown in Fig.~\ref{fig:phi_vs_psi_0_5pi}(b), numerical diagonalization of $\hat{G}$ yields a spectrum of eigenvalues $\lambda$ that have no appreciable real parts, allowing us to conclude that the state $\psi$ is dynamically stable.
The state $\varphi$, having large overlap with $\psi$, may be expected likewise not to possess any exponentially growing modes in its dynamical evolution. To check if this is the case, we directly propagate Eqs.~\eqref{eq:free} starting from the state $\varphi$ resulting from the optical approach. Calculations of dynamics for an interval of 1 second, shown in Fig.~\ref{fig:dynamics_0_5pi}, reveal that the DB soliton stays in place, corroborating its dynamical robustness after being created from our experimentally tractable optical pulse-based procedure. On a finer scale, both dark and bright 
solitons undergo breathing dynamics, which is the result of the state $\varphi$ not being identical to the stationary configuration $\psi$ initially, but the solitons nevertheless remain localized at their initial positions.

\section{Dark--bright soliton lattices}

\begin{figure}
\begin{centering}
\includegraphics{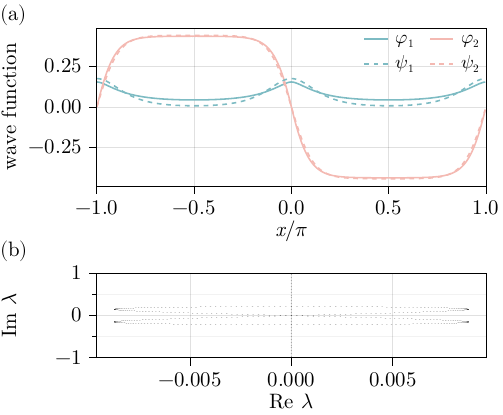}
\par\end{centering}
\caption{Stationary state of the system interacting with light fields for $x\in[-\pi, \pi)$ under periodic boundary conditions, and the BdG analysis for an infinite lattice.
Nonlinear coupling ratios corresponding to \textsuperscript{87}Rb were used. (a) Solid lines: stationary solution of Eqs.~\eqref{eq:full-gpe} representing the lowest-energy state of the dark-state manifold. The wave function $\varphi_3$ is not shown since the population of this state is less than $10^{-5}$. 
Dashed lines: corresponding stationary solutions of Eqs.~\eqref{eq:free}.
(b) BdG stability spectrum for the case of an infinite periodic soliton lattice,
whose single cell is shown in (a). The purely imaginary eigenvalues span the whole of the imaginary axis, but the view is restricted to the interval $[-1, 1]$.
The displayed result was obtained using the Floquet--Hill approach by sampling 101 values of the quasimomentum-like parameter $q$.
}\label{fig:phi_vs_psi_1pi}
\end{figure}

Next we extend our considerations to dark--bright soliton lattices by enlarging the
extent of the $x$-axis. The interval $x\in[-\pi,\pi)$ allows us
to study a single cell of a lattice, as shown in Fig.~\ref{fig:phi_vs_psi_1pi}(a).
Here we display a comparison of $\varphi$, representing the stationary
state of Eqs.~\eqref{eq:full-gpe}, with the stationary state $\psi$
of the free system, obtained under the constraint \eqref{eq:constraint}.
As in the case of a single dark--bright soliton, the overlap between
the dark solitons is strong, yielding a total overlap $\braket{\varphi}{\psi}=0.9975$.
Calculating the BdG spectrum we again find that the system is dynamically
stable. This is in line with previous studies that demonstrated
the existence of an exact stationary solution of Eq.~\eqref{eq:free}
in the form of a pair of DB solitons. It has been also
shown that these states are stable in certain
regimes of nonlinear couplings \citep{Katsimiga2018}.

Having obtained the solution $\psi$
for a single cell, we can study the stability of an infinite periodic
lattice using the Floquet--Hill approach~\cite{DeconinckKutz2006}. 
For a periodic state $\psi$,
the $\hat{G}$ operator in Eq.~\eqref{eq:BdG} becomes periodic, allowing
us to focus on a single cell of a lattice and introduce a quasimomentum-like
parameter $q$. The linearization equation then transforms into $\hat{G}_{q}(x)B_{q}(x)=\i\lambda_{q}B_{q}(x)$
upon introducing eigenvectors of the form $B_{q}(x)=\e^{\i qx}C_{q}(x)$.
Constructing $\hat{G}_{q}$ and diagonalizing for a range of values $q\in[-\pi/L,\pi/L)$
provides us with stability signatures for an infinite lattice. This
time, apart from a range of purely imaginary eigenvalues $\lambda_q$
we obtain a host of eigenvalues possessing non-zero real parts, as
shown in Fig.~\ref{fig:phi_vs_psi_1pi}(b). 
The values enable us
to estimate the time before the instability has grown appreciably.
The largest eigenvalue of $0.0087$ leads to the time interval of
$1/\lambda_q=\unit[4.8]{ms}$.
The stability spectrum calculated for the Manakov case (equal nonlinear couplings) looks identical to the one shown in Fig.~\ref{fig:phi_vs_psi_1pi}(b). However, the stability of soliton lattices  does differ in the two cases, as we discuss next.

\subsection{Manakov case}

\begin{figure}
\begin{centering}
\includegraphics{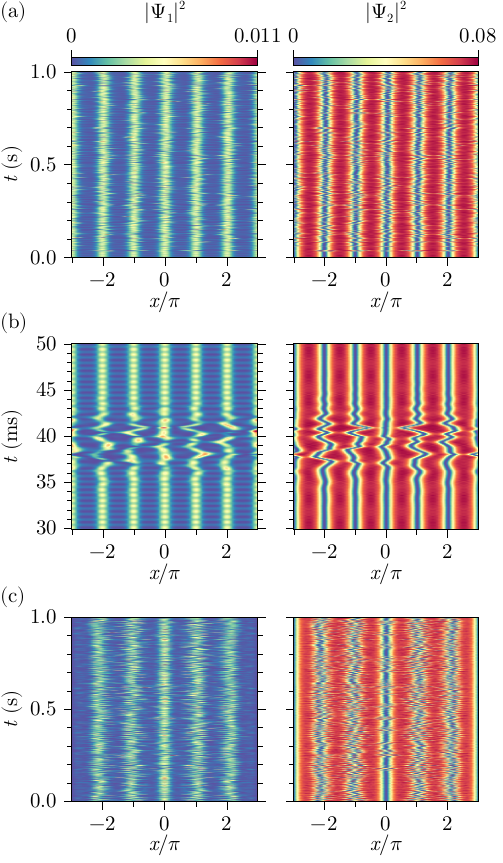}
\par\end{centering}
\caption{Manakov case: evolution of
the densities $|\Psi_{1}|^2$ and $|\Psi_{2}|^2$,
governed by Eqs.~\eqref{eq:free} and assuming $g_{11}=g_{12}=g_{22}=100$, starting from $\varphi_{1}$ and $\varphi_{2}$. (a) Calculation for a time interval of $\unit[1]{s}$, periodic boundary conditions. (b) A close-up view of the dynamics around the first instability in (a). (c) Same as (a) for hard-wall boundary conditions.}\label{fig:dynamics_3pi}
\end{figure}

First, let us consider the dynamics of larger lattices in the case of equal nonlinear couplings. To this end we set $g_{11}=g_{12}=g_{22}=100$ and
calculate the dynamics
governed by Eqs.~\eqref{eq:free} for a lattice on the interval $x\in[-3\pi,3\pi)$,
starting from state $\varphi$, i.e., with the initial condition
being $\Psi_{i}(t=0,x)=\varphi_{i}(x)$. We note that while the stability
analysis is only applicable to the stationary state $\psi$, the large
overlap between $\varphi$ and $\psi$ leads to a similar evolution
of these states;
the only essential difference is that states evolving from $\varphi$ exhibit breathing.
The evolution of $|\Psi_{1}|^{2}$ and $|\Psi_{2}|^{2}$
is displayed in Fig.~\ref{fig:dynamics_3pi}(a). The instability sets
in after approximately $\unit[35]{ms}$ and results in solitons starting
to oscillate around their initial positions, as can be seen in the
close-up view in Fig.~\ref{fig:dynamics_3pi}(b).
However, as we can see, the system recurs,
and the densities of the two components return to those resembling the initial configuration.
This can be intuitively understood on the basis of the integrable
(and hence generically quasi-periodic) dynamics of the Manakov
model of Eqs.~(\ref{eq:free}) for the case of equal $g_{ij}$.

In our simulations the instability develops considerably later compared to the BdG prediction of 4.8\,ms. This reflects the fact that the magnitude of the overlap $\braket{\psi}{u_i}$ between the initial state $\psi$ and the unstable modes $u_i$ is of order $10^{-11}$, therefore, it takes a considerable time for the unstable modes to grow \footnote{For example, the mode $\delta\psi(x)=a(x)\e^{\lambda t}$ with a magnitude of $a(x_0)=10^{-11}$ at some point $x_0$ will reach the value of $10^{-1}$ at $x_0$ in time $10\ln(10)/\lambda\approx23/\lambda$ (rather than $1/\lambda$).}.
Generally, the exact time at which the instability manifests, as well as quantitative features of the subsequent evolution, depend on the accuracy of preparation of the initial state.

To confirm that the instability appearing in the calculated dynamics is indeed the result of the growth of the unstable modes, we calculate the BdG spectrum for a smaller lattice spanning $x\in[-2\pi,2\pi)$ and find that it possesses a single unstable mode. Then, calculating the evolution for the fields-free eigenstate $\psi$ and tracking the quantity
\begin{equation}
\Delta\Psi(t)=\int\d x\,\left||\Psi(t,x)|^{2}-|\Psi(0,x)|^{2}\right|
\end{equation}
[where $\Psi(0,x)=\psi(x)$] we confirm that it grows as $\e^{\lambda t}$ with $\lambda$ coinciding (at an accuracy of three significant digits) with the real part of the eigenvalue of the unstable BdG mode.

While still considering the Manakov case,
let us assess the stability of the lattice in the presence of a trapping potential. As is well known \cite{Busch2001}, in the presence of a harmonic trap, a single soliton will oscillate harmonically around the center of the trap. Therefore, the lattice can be sustained only under very specific conditions, where the trap frequency and the distance between the solitons are fine-tuned such that the interaction between the solitons cancels the effect of the trap. Clearly, an optical-box trap \cite{Gaunt2013}, allowing for a homogeneous condensate, is more appropriate for the lattice considered herein, especially if its length is chosen in integer multiples of the spatial period of the dark-state potential.
In the hard-wall case, we could not converge to a stationary state $\psi$ of the soliton-lattice form in the \emph{fields-free} case. Therefore, the BdG analysis is inapplicable, and the stability of the lattice can only be determined by a direct propagation of the initial state $\varphi$. The latter could be successfully obtained using the Newton--Raphson iteration, exactly as in the periodic case.
The simulations of dynamics of a system subject to hard-wall boundary conditions are shown in Fig.~\ref{fig:dynamics_3pi}(c). Contrary to the case of periodic boundary conditions, the solitons never return to a configuration similar to the initial one.
Nevertheless, the scattering of the solitons does not destroy the overall lattice structure.

\begin{figure}
\begin{centering}
\includegraphics{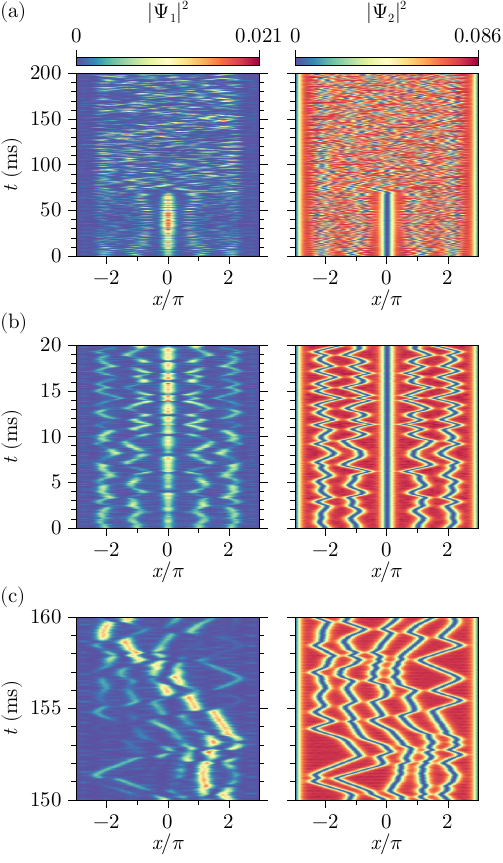}
\par\end{centering}
\caption{The case of \textsuperscript{87}Rb: evolution of the densities $|\Psi_{1}|^2$ and $|\Psi_{2}|^2$, governed by Eqs.~\eqref{eq:free} with hard-wall boundary conditions, starting from $\varphi_{1}$ and $\varphi_{2}$. Nonlinear coupling ratios corresponding to \textsuperscript{87}Rb were used. (a) Dynamics up to $t=\unit[200]{ms}$. (b) A close-up view of the dynamics for the first $\unit[20]{ms}$. (c) A close-up view of the dynamics around $t=\unit[155]{ms}$.}\label{fig:dynamics_Rb-87}
\end{figure}

\subsection{Case of \textsuperscript{87}Rb}

Now let us turn to the case of unequal nonlinear couplings. First, we remark that the ability of the system to recur in the periodic case, as observed for the Manakov system [cf. Fig.~\ref{fig:dynamics_3pi}(b)], is related to the integrability of the system and hence is very sensitive to the Manakov condition $g_{11}=g_{12}=g_{22}$. For example, if we change any one of the couplings to the value of 99 while keeping the other two at the value of 100, the recurrence is no longer observed. Instead, the scattering of the solitons becomes more chaotic; once the instability sets in, the dynamics becomes similar to that of the hard-wall case shown in Fig.~\ref{fig:dynamics_3pi}(c). The time for the instability to set in, however, does not significantly change.

To be more specific, we now discuss the results obtained for nonlinear couplings corresponding to \textsuperscript{87}Rb. The results for the gas confined in a box-shaped trap are shown in Fig.~\ref{fig:dynamics_Rb-87}. Figure \ref{fig:dynamics_Rb-87}(a) shows the dynamics calculated up to 200\,ms after the sudden quench of the optical fields, starting from the stationary state $\varphi$ obtained using our optical scheme. As we can see, the solitons move far away from their initial positions, and the regular lattice structure gets destroyed on a timescale of $\sim\unit[10]{ms}$. Due to the symmetry of the system, the central soliton remains in place for a longer period of time. However, while its dark component exhibits no appreciable change of density, the bright one experiences the transfer of population to the other bright solitons and back. This is demonstrated in a close-up view of the initial 20\,ms of dynamics provided in Fig.~\ref{fig:dynamics_Rb-87}(b). It is apparent that all the solitons, except for the central one, start moving away from their initial positions right from the beginning of the evolution. In the long run, the scattering of bright solitons results in the formation of a single pronounced bright soliton, displayed in Fig.~\ref{fig:dynamics_Rb-87}(c) showing the evolution around the $t=\unit[155]{ms}$ mark. Meanwhile, the dark solitons retain their separate characters.

\section{Conclusions and Future Challenges}

In summary, we have shown that a 3-component BEC interacting with
the laser fields in the $\Lambda$ configuration can be used to create a reproducible dark--bright soliton lattice, which is the lowest-energy
state of the dark-state manifold.
In most of our calculations we used realistic scattering length ratios corresponding to \textsuperscript{87}Rb. First, we focused on the solution of the form of a single DB soliton  (under Neumann boundary conditions). We found that when the optical fields are quenched, such a DB soliton retains its form on the timescale of the lifetime of the condensate. This stability is attributed to the fact that the resulting state $\varphi$ is close to being a stationary state $\psi$ of the fields-free system. The fact that $\varphi$ is not exactly a stationary state leads to breathing dynamics, but otherwise the state evolves as a stationary state would. Similar results were obtained for the case of two DB solitons under periodic boundary conditions. Next, we studied the stability of soliton lattices. The Floquet--Hill variant of the BdG analysis revealed that relevant infinite periodic lattices of DB solitons that are stationary states of the fields-free system are dynamically unstable. The instability causes oscillatory deviations of the solitons away from their initial positions. The dynamics of the DB soliton lattices prepared using our optical scheme exhibits a similar instability. As a result, the lattice remains stable  only for a few tens of milliseconds after the quench of the optical fields. This is so both for a periodic lattice and one confined in a box potential.

Notably, we also found that in the case of equal nonlinear couplings, a periodic lattice exhibits recurrence, returning to the initial configuration and thus retaining the lattice form on the order of 1\,s. However, a one-percent deviation from exact equality of the couplings already suppresses the recurrence. The departure of ratios of $g_{ij}$ from unity leads to a breaking of integrability, which imparts chaotic dynamics. Although the rate of the exponential growth of the instability is only weakly dependent on the nonlinear coupling ratios as we depart from the Manakov limit, the eventual fate of the system is drastically affected by the integrable or non-integrable scenario and results in recurrence in the former and in rapid dissolution of the soliton lattice in the latter.

As a possible extension of the present work, it would be particularly interesting to explore similar phenomena in 
systems involving a larger number of components to investigate whether more
exotic soliton configurations can be accordingly prepared. Moreover, 
it would be natural to extend relevant considerations to higher dimensional
settings and the highly active investigation of vortex--bright solitary
waves~\cite{richaud1,richaud2,ohsawa2026smallmassasymptoticsmassivepoint}.

\begin{acknowledgments}
This project was supported by the Research Council of Lithuania (RCL)
grant No.~S-LJB-24-2 and the JSPS Bilateral Program No.~JPJSBP120244202.
This work has also received funding from COST Action POLYTOPO CA23134,
supported by COST (European Cooperation in Science and Technology). 
This research was supported by the U.S. National Science Foundation under the award PHY-2408988 (PGK). 
This research was partly conducted while P.G.K. was  visiting the Okinawa Institute of Science and
Technology (OIST) through the Theoretical Sciences Visiting Program (TSVP), the University of
Sydney through the visitor program of the Sydney Mathematical Research Institute (SMRI) and the Department of Mechanical Engineering at Seoul National 
University through a Fulbright Fellowship. Their support is gratefully acknowledged.
Finally, this work was also  supported by a grant from the Simons Foundation [SFI-MPS-SFM-00011048, P.G.K]. 
\end{acknowledgments}

\section*{Data Availability}

The data that support the findings of this study were generated by numerical simulations using the QuantumHamiltonians.jl package \cite{QuantumHamiltonians.jl} developed in Julia \cite{Julia} and depending on a number of numerical software packages \cite{JuliaDiffEq,JuliaETDRK,JuliaNonlinearSolve,JuliaArnoldiMethod,JuliaFFTW,JuliaKrylovKit}. The figures in the text have been produced using Makie.jl \cite{Makie.jl}.

\end{document}